\documentclass[12pt,a4paper,twocolumn]{article}
\usepackage[left=2cm,right=2cm,top=2cm,bottom=2cm]{geometry}
\usepackage{authblk}
\usepackage{cite}

\usepackage[nottoc,other]{tocbibind}

\usepackage{sectsty}
\allsectionsfont{\centering}
\sectionfont{\fontsize{10}{15}\selectfont}
\subsectionfont{\fontsize{9}{12}\selectfont}
\subsubsectionfont{\fontsize{9}{12}\selectfont}
\usepackage[titletoc]{appendix}
\usepackage{listings,listingsutf8}
\usepackage{lmodern}
\usepackage{xcolor}
\usepackage{relsize} 
\usepackage{comment}
\usepackage{graphicx}
\graphicspath{ {images/} }
\usepackage{bm}
\usepackage[centerlast,scriptsize]{caption}
\usepackage{subcaption}
\usepackage{floatrow}
\usepackage{mathtools}
\usepackage{float}
\usepackage{amsfonts,amsmath,amssymb,amsthm}
\usepackage[hidelinks,colorlinks=false]{hyperref}
\usepackage[capitalise]{cleveref}

\usepackage{physics}
\usepackage{siunitx}
\usepackage{setspace}
\usepackage{ragged2e}
\usepackage{pdfpages}
\usepackage{multicol}

\usepackage{fancyhdr}

\title{\textbf{Cycle and bit accurate chaos-based communication system\footnote{Chaos-based spread-spectrum communication system, in \textbf{Part C}}}}
\author[1]{\small \textbf{Christian Nwachioma}\thanks{christian.nwachioma@gmail.com}}
\affil[1]{\textbf{ CIDETEC, Instituto Polit\'ecnico Nacional, Mexico City 07700, Mexico} }
\providecommand{\keywords}[1]{\textbf{\textit{Index terms --- }} #1}
\date{}
\begin{document} 
\onecolumn  
\maketitle 	
\thispagestyle{fancy}
\normalsize
\begin{abstract}
	\noindent Adaptive control is a control method that has an adaptation mechanism that reacts to model uncertainties. The control method is used to realized synchronization of a new chaotic system in a unidirectional master-slave topology. The master chaotic system and the slave system are adopted as transmitter and receiver, respectively for the purpose of secure communications. Both analog and digital designs are realized. The digital system is a cycle and bit accurate design having a system rate of $450MHz$. The design is targeted at an \textit{Artix-7 Nexys 4} FPGA. The transmitters are accordingly modulated by analog signals and fixed-point signals of different resolutions, sampling and frequencies. Although the adaptive controller tends to react on introduction of a modulating signal, we show that with a special detection mechanism or filtering including exponential smoothing, the choice of which depends on the nature of the modulating signal, it is possible to recover the modulating signal at the receiver. Moreover, chaos-based spectrum-spectrum communication systems are realized based on adaptive synchronization of the new chaotic system. Furthermore, in order to ascertain the robustness of the adaptive controller, the modulated signal is transmitted via an \textit{awgn} channel and the probability of error and bit-error-rate (BER) computed by sweeping across sets of SNR and noise power values in a Monte Carlo simulation. It turns out that the probabilities of error or error rates are reasonably low for effective communication through mediums with certain noise conditions. Hence, the adaptive controlled communication system can be robust.	
	\noindent\newline\newline\keywords{ Chaotic system; secure communications; adaptive control; analog and digital signals}
\end{abstract}
\begin{multicols}{2}
	\begin{center}
		\section{INTRODUCTION}
	\end{center}
	\label{secdigital}
	The communication system realized in \textit{Part A} is based on floating-point arithmetic. While that approach can accurately represent real-world values and prevents overflows, its implementation utilizes more resources such as high-power, memory and cost. Besides, it can be deemed out of touch with contemporary communication systems. To eschew this situation, a trade-off can be made by realizing a fixed-point design. In order to demonstrate the applicability of the new chaotic system in digital communication, we realize a cycle and bit accurate design using the \textit{Vivado System Generator} (VSG). The digital chaotic signals will be modulated by binary signals and higher resolutions signals of different sampling rates and frequencies. Furthermore, states of the digital chaotic transmitter and the synchronized receiver will be exploited for spread-spectrum communication application in \textit{Part C}. 
\end{multicols}
\begin{multicols}{2}
\begin{center}
		\section{TRANSMITTER \& RECEIVER}
\end{center}
	\label{txdigital}
	In this section, we realize a fixed-point design which is a necessary first step for digitization of the new chaotic system. Floating point arithmetic may also be used especially for a processor that has the \textit{floating-point unit} (FPU), but the fixed-point design is more economical in terms of hardware resource utilization. In realizing a fixed-point design, it is important to use a sample rate and amplitude resolution that can satisfactorily and finely represent the signal of interest. Hardware like FPGAs or other embedded platforms manipulate the fixed-point arithmetic in streams of \textit{zeros} and \textit{ones}. Streams of \textit{zeros} and \textit{ones} derived from a chaotic transmitter can be utilized for the purpose of digital communication security since chaotic signals are inherently broadband and noise-like yet deterministic. 
	
\begin{figure}[h!]
		\includegraphics[width=0.9\linewidth]{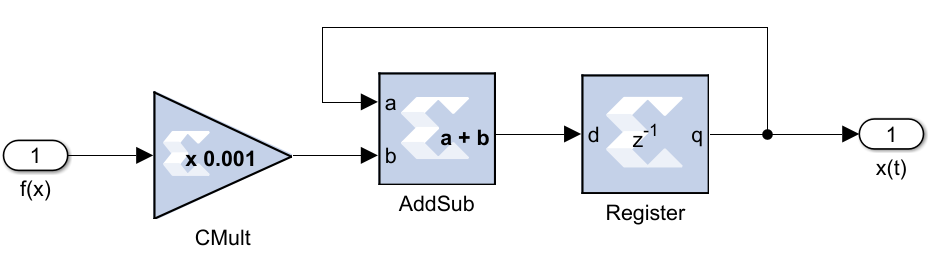}
		\caption{Euler integration method realized on VSG.}
		\label{integ}
	\end{figure}

\begin{figure}[h!]
		\includegraphics[width=0.95\linewidth]{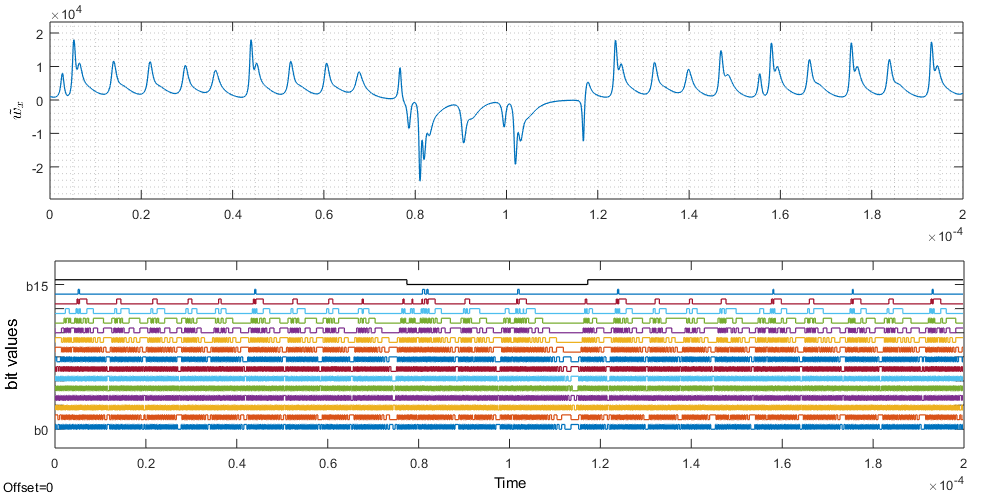}
		\caption{\textit{Top}: Sampled and quantized $w_x$ chaotic signal with $16$-bit resolution. \textit{Bottom}: digitized $w_x$ chaotic signal}
		\label{digitizedwx}
	\end{figure}

	For the sake of referencing, let us define $\tilde{w} := (\tilde{w}_x, \tilde{w}_y, \tilde{w}_z)$ corresponding to $w = (w_x,w_y,w_z)$ of the analog design case in \textit{Part A}, to be a vector of fixed-point signals describing states of the digital transmitter. Where no confusion can arise, $\tilde{w}$ will also refer to the digitized equivalence of the fixed-point signal. Using the VSG, we can realize a cycle and bit accurate design of the new chaotic transmitter. For a demonstration of the design, we choose a sampling frequency of $450MHz$. The motivation for choosing this speed is partly because in a chaos-based spread-spectrum application using FM technology to be presented later in \textit{Part C}, the system rate has to be greater than twice the carrier frequency\cite{shannon1984communication,wyner1998introduction} and we shall be using carrier frequencies up to about $200MHz$. 
	Moving forward, the quantization is unbiased and rounded capturing both positive and negative values in $16$ bits resolution. The design uses Euler integration method, which is modeled as shown in \cref{integ}. The step size equals $0.001$. For the chosen resolution, a common scaling factor $s_f = 3107$ will suffice for the communication system. Based on this scaling factor, the initial condition in fixed-point for the transmitter is $\tilde{w}(0) = (1032, -3107, 0)$. Dividing by the scaling factor $s_f$, it can be seen that the initial condition is approximately equivalent to the analog design case in \textit{Part A} with the little difference arising due to sampling and quantization. At the level of hardware, the transmitter is digitized and have the initial condition: $\tilde{w}_x(0) = `1000010000001000'$, $\tilde{w}_y(0) = `0000110000100011'$ and $\tilde{w}_z(0) = `1000000000000000'$. 
	In \cref{digitizedwx}, is a transmitter signal $\tilde{w}_x$, sampled and quantized at $16$-bit resolution and the corresponding $16$-level binary equivalence. 
	The most significant bit (MSB) is $b_{15}$ and it is the sign bit. A logic \textit{high} for the sign bit represents a positive (including zero). A logic \textit{low} represents a negative. This can be correlated with the sampled and quantized version at the specific sample time. The weight of the bits decreases downward and the least significant bit (LSB) is $b_0$. As expected, the transition rate between \textit{low} and \textit{high} logic states is faster while traversing downward. This patternless sequence of bits can be used to spread or encode another sequence of bits which has a slower rate and usually a shorter word-length.
	
\begin{figure}[h!]
		\includegraphics[width=0.95\linewidth]{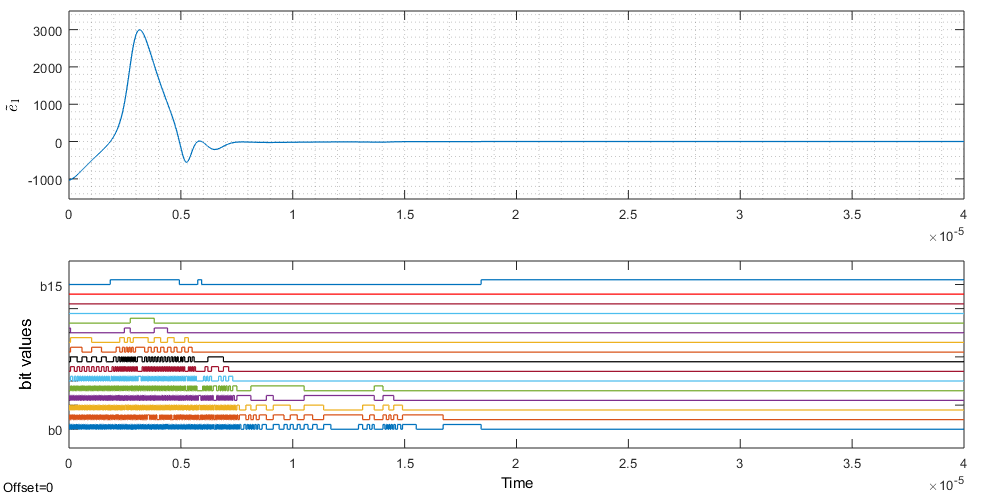}
		\caption{\textit{Top}: discrete-time quantize-amplitude synchronization error with $16$-bit resolution. \textit{Bottom}: $16$ bits digitized version of the synchronization error.}
		\label{digsync_error}
	\end{figure}

\begin{figure}[h!]
		\includegraphics[width=0.8\linewidth]{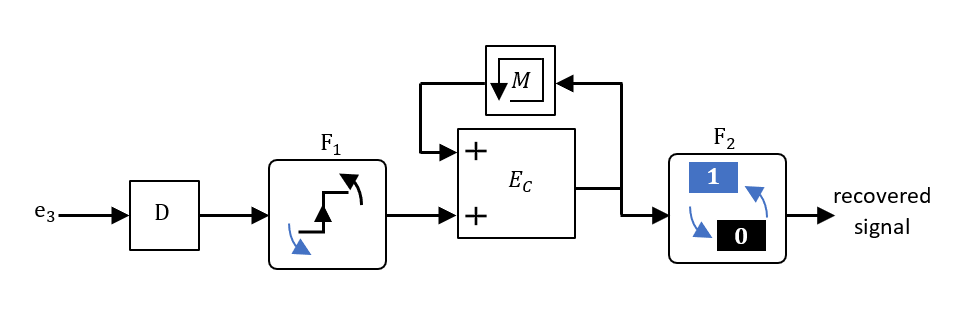}
		\caption{Detection mechanism at the receiver for a binary information signal}
		\label{sg_detect}
	\end{figure}

\begin{figure}[h!]
		\includegraphics[width=0.5\linewidth]{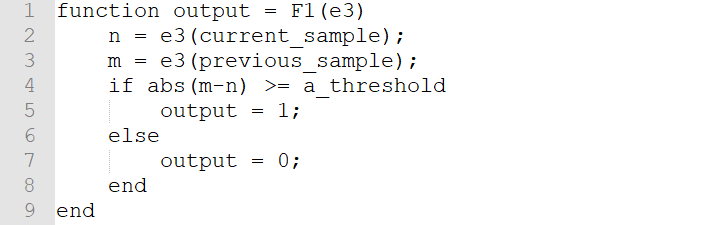}
		\caption{The decision function for realizing edge detection.}
		\label{decision_fcn}
	\end{figure}

\begin{figure}[h!]
		\includegraphics[width=0.5\linewidth]{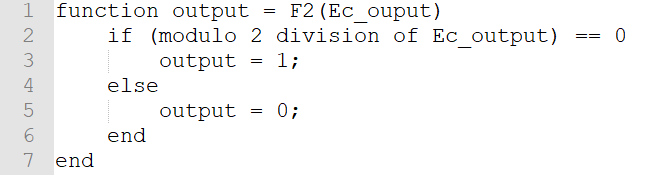}
		\caption{Assignment of binary states based on number of detected edges.}
		\label{decision_fcn2}
	\end{figure}

\begin{figure}[h!]
		\includegraphics[width=0.95\linewidth]{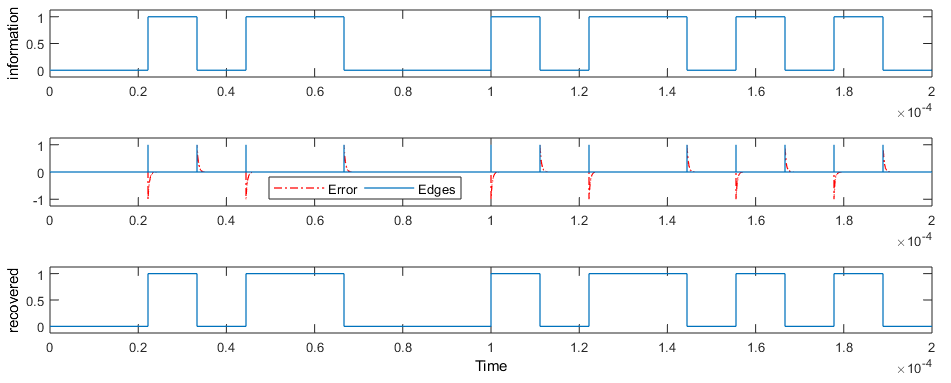}
		\caption{Transmission and recovery of a binary signal through the digital chaotic transmitter-receiver network in the absence of channel noise.}
		\label{sg_sqwv}
	\end{figure}

	Now, let $\tilde{\sigma}:= (\tilde{\sigma}_x, \tilde{\sigma}_y, \tilde{\sigma}_z)$ corresponding to $\sigma = (\sigma_{x},\sigma_{y},\sigma_{z})$ of the analog design case in \textit{Part A}, be a fixed-point vector of states of the receiver. Where no confusion can arise, $\tilde{\sigma}$ will also refer to the digitized equivalence of the fixed-point values. Furthermore, $\tilde{u}_1$, $\tilde{u}_2$ and $\tilde{u}_3$ can be used to reference the fixed-point or digital equivalences of the control inputs. Since, we are dealing with digital designs that can be implemented on FPGAs or other embedded platforms, we can assume that corresponding transmitter and receiver coefficients are identical. However, the initial conditions can differ since the transmitter and receiver are remotely located, started at different times and time lags can exist during system initialization or transmission. With the positive gains $k_1 = 2s_f$, $k_2 = s_f$, $k_3 = 3s_f$, initial condition $\tilde{\sigma}(0) = (0, -4660, 1553)$ and a fixed-point realization of the adaptive controller in \textit{Part A}, synchronization can be realized between the digital transmitter and receiver (see \cref{digsync_error}). The result is from a transmitter that is \textit{unmodulated} and in this situation, synchronization is easy to realize. Moreover, since VSG is cylce and bit accurate, the \textit{Vivado integrated logic analyzer} (ILA) can be used to view corresponding digital signals provided the following implementation stages are passed: VHDL or Verilog generation, \textit{synthesis}, \textit{place and route}, bit-stream generation and download into an appropriate FPGA. In the present case, we are combining \textit{VSG} and \textit{MATLAB-SIMULINK} to display the corresponding system-level digital signals. Implementation of the chaotic transmitter on a \textit{Nexys 4 Artix-7} FPGA is reported elsewhere.
	\end{multicols}
\begin{multicols}{2}
	Next, we exploit the unpredictable nature of the chaotic digital transmitter to securely transmit discrete-time signals of different samplings, resolutions and frequencies. The broadband and noise-like nature of the chaotic signal hides the information signal so that even though the encoded signal is intercepted, the interceptor cannot detect the presence of meaningful signal within. The modulated transmitter signal is used in the controller to produce digital control signals which are subsequently used in attempting to drive the receiver to track the transmitter. However, in this situation, the receiver cannot track the transmitter perfectly as the modulation contains a ``\textit{foreign}'' signal, which produces an error. The error is usually not straightforwardly commensurable to the information signal due to action of the adaptive controller. Hence, an appropriate detection mechanism may have to be applied in order to recover the original information signal. 
	Footprints of error due to the information signal are expected since the digital \textit{integrators} in the adaption mechanism tries to compensate for any fluctuations.
	\begin{center}
		\subsection{NUMERICAL EXPERIMENT}
		\label{exponebit}
		\subsubsection*{1-BIT SIGNAL TRANSMISSION}
	\end{center} 
	Here, we are transmitting binary states of \textit{zeros} and \textit{ones} through the digital communication system. Initial conditions and coefficients of the transmitter and receiver are configured as before. The binary information signal has a rate of $1MHz$, that is, $450$ times slower than the system rate. Unlike the case in \textit{Part A}, the process of detecting the information signal is rather more involving but can be fine-tuned to be more accurate. \cref{sg_detect} shows the information detection mechanism. Since the information signal is used to modulate $\tilde{w}_z$ chaotic transmitter signal, the difference with the corresponding receiver signal $\tilde{\sigma_{z}}$ can be used to regenerate the information signal. Therefore, input to the mechanism is the error $e_3$, which is delayed appropriately by the element $D$ in order to allow for synchronization of the free transmitter and controlled receiver. Next, the delayed error signal is subjected to a decision mechanism $F_1$. The role of $F_1$ is given in the snippet of code in \cref{decision_fcn}. Provided the threshold is properly chosen, $F_1$ can detect points in time where the information signal transitions from one state to the next. $F_1$ places a marker (or an edge) at every point of the transition. Essentially, $F_1$ is an edge detector that allows us to specify a sensitivity threshold\footnote{Strictly speaking, the edge is not really a vertical line, but a rectangle of width 1 sample time.}. The next unit $E_c$ is a counter that counts the number of edges detected and the unit labeled $M$ is a memory/delay device use to preempt algebraic loop in the feedback path. Finally, there is the unit labeled $F_2$ which performs a modulo $2$ division on the number of edges as they come and assigns a \textit{zero} or \textit{one} according to the code snippet in \cref{decision_fcn2}. Care is taken to use only hardware-friendly codes and model blocks. Setting the variable ${a\_threshold}=0.5$ in the $F_1$ function snippet, the result of the experiment is shown in \cref{sg_sqwv}. It can be seen that the transmitter-receiver network, the adaptive controller and the detection mechanism are able to provide a successful communication in this particular case of transmitting binary information signal.
\end{multicols}
	
\begin{figure}[h!]
		\begin{subfigure}[b]{\linewidth}
			\begin{subfigure}[b]{0.5\textwidth}
				\includegraphics[width=\linewidth]{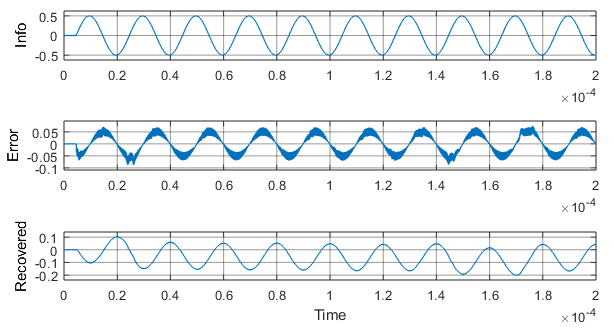}
				\caption{Input rate: $4.5MHz$. Signal property: $16$-bit resolution, $50kHz$.}
			\end{subfigure}
			\hfill
			\begin{subfigure}[b]{0.5\textwidth}
				\includegraphics[width=\linewidth]{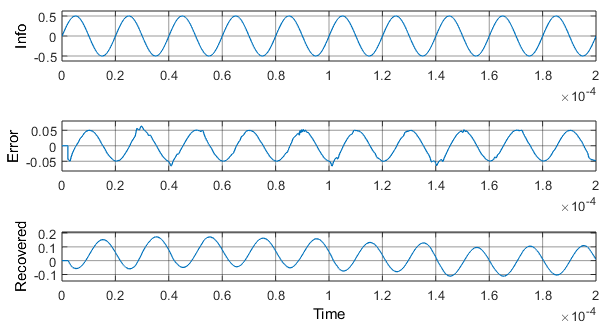}
				\caption{Input rate: $450MHz$. Signal property: $16$-bit resolution, $50kHz$.}
			\end{subfigure}
		\end{subfigure}
		\hfill	
		\begin{subfigure}[b]{\linewidth}
			\begin{subfigure}[b]{0.5\textwidth}
				\includegraphics[width=\linewidth]{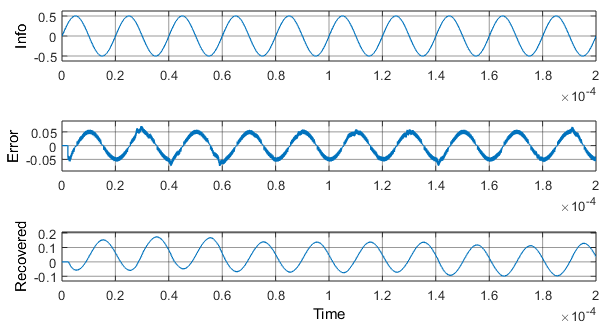}
				\caption{Input rate: $450MHz$. Signal property: $8$-bit resolution, $50kHz$.}
			\end{subfigure}
			\hfill
			\begin{subfigure}[b]{0.5\textwidth}
				\includegraphics[width=\linewidth]{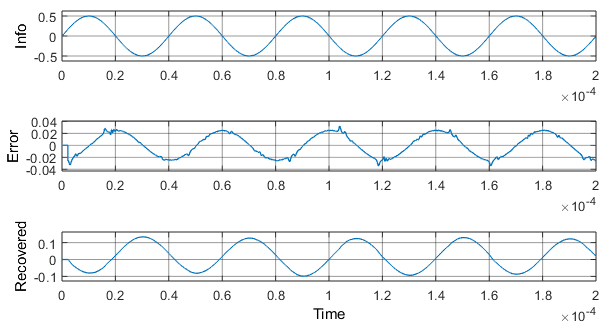}
				\caption{Input rate: $450MHz$. Signal property: $16$-bit resolution, $25kHz$.}
			\end{subfigure}
		\end{subfigure}	
		\caption{Transmission and reception of information with different sampling rates, signal resolutions and frequencies.}
		\label{sampled_st}
	\end{figure}

\begin{figure}[h!]
	\includegraphics[width=0.95\linewidth]{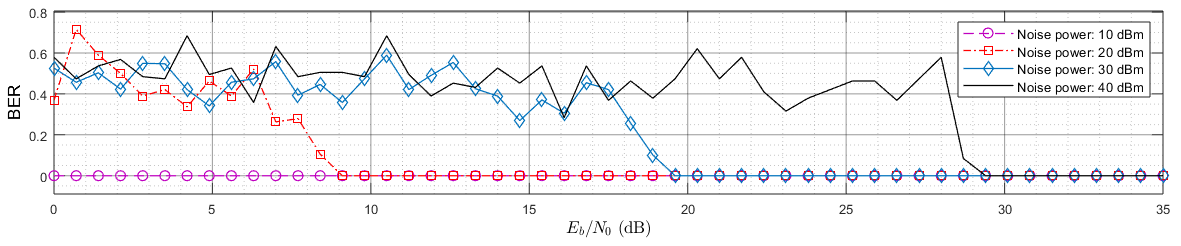}
	\caption{BER vs SNR in $E_b/N_0$ mode.}
	\label{EBN0}
\end{figure}

\begin{figure}[h!]
	\includegraphics[width=\linewidth]{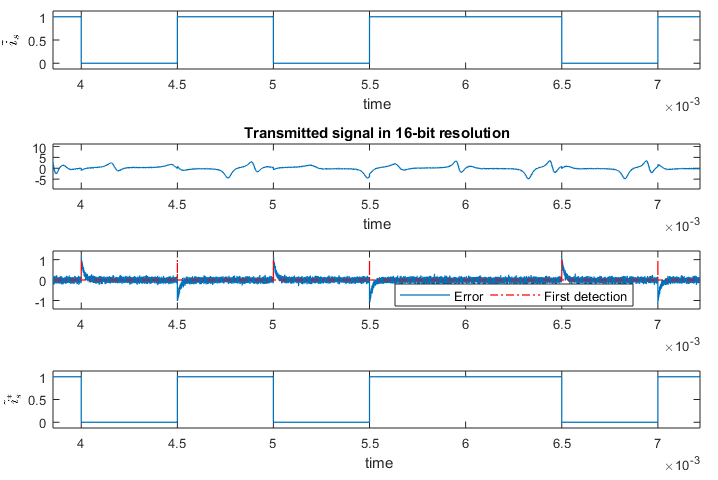}
	\caption{Communication in the presence of a channel noise of $20\enspace dB$ in $E_b/N_0$ mode.}
	\label{sg_sqwv_20db}
\end{figure}

\begin{multicols}{2}
	\begin{center}
		\subsection*{HIGHER RESOLUTION SIGNAL TRANSMISSION}
	\end{center}
	In the previous numerical experiment, the transmitted signal was binary having only two distinct states of \textit{zeros} and \textit{ones}. Here, we deem it fit to experiment with modulating signals having different resolutions, samplings and frequencies. We note the effects of these quantities on the recovered information signal. We assume that the modulating signal is a sine wave with amplitude of $0.5$ and zero phased. The transmitter and receiver parameters are set as before.
	Referring to \cref{sampled_st}, first, we consider the modulating signal to be of $16$-bit resolution and $50kHz$ and sampled at $4.5MHz$.
	For comparison of sampling effect, the experiment is repeated with the signal having the same properties but sampled at the system rate of $450MHz$. It can be seen that a faster sampling of the modulating signal reduces the reception error. Furthermore, assuming the modulating signal has the properties, $8$-bit resolution and $50kHz$ and sampled at $450MHz$, the effect of signal resolution, compared with previous case of $16$-bit resolution, can be seen in the figure. Also, assuming the information signal has the properties, $16$-bit resolution and $25kHz$ and sampled at $450MHz$, the effect of signal's frequency, in comparison to the previous case of $50kHz$, can be appreciated in the figure. It is observed that with a finer resolution (or sampling),  the reception error becomes lesser and the need for filtering might not arise. It is also observed that with an increase in frequency of the modulating signal, the amplitude of the recovered signal approaches the amplitude of the original information signal. Above a certain threshold of frequency, the amplitude of the recovered signal equals the amplitude of the original information signal such that no amplification is necessary. However, it is noted that frequency of the modulating signal should not exceed the sampling frequency. Although, the sampling frequency can be increased, the increase is subject to hardware limitations. For instance, in theory, we can increase the sample rate of the system very much above $450MHz$ but FPGAs and other embedded platforms as of today, have maximums in the neighborhood of $500MHz$.
\end{multicols} 
\begin{multicols}{2}
	\begin{center}
		\section{PERFORMANCE ANALYSIS}
	\end{center}
In the previous sections, the numerical experiments were conducted in an ideal environment. That is, the communication channels were devoid of noise or disturbances. Now, we shall introduce the white Gaussian noise of significant power level into the communication channel in order to test the robustness of the adaptive controller and detection mechanisms or filters.
\end{multicols}
\begin{multicols}{2}
\begin{center}
	\subsection{BIT ERROR RATE}
\end{center}
Bit error rate (BER) is the ratio of bit error to the total number of bits received in a transmission. We are repeating the experiment in \cref{exponebit} but this time, we are adding a channel noise. The signal to noise ratio (SNR) in energy per bit to noise power spectral density ($E_b/N_0$), is varied from $0$ to $35~dB$. Within the range of SNR, four different noise power levels viz. $10~dBm$, $20~dBm$, $30~dBm$ and $40~dBm$ are considered. The transmitter, controller and receiver are set as in \cref{exponebit}. A Monte Carlo simulation is performed with the SNR updated in every run. The BER is collected at the end of every run and graphed against the corresponding SNR as shown in \cref{EBN0}. For the noise power of $10~dBm$, the BER is zero within the specified SNR range. For higher noise power levels, it can be seen that the SNR at which the BER goes to zero increases with increasing noise power. \cref{sg_sqwv_20db} shows a transmitted signal with an SNR of $20~dB$ of which the noise power is $30~dBm$. It can be seen that the adaptive controller and the detection mechanism is able to faithfully recover the information. 
\end{multicols}
\begin{multicols}{2}
	\begin{center}
		\section{CONCLUSION}
	\end{center}

Adaptive control is a control method that has an adaptation mechanism that reacts to model uncertainties. While this feature might be desirable for stability of systems like aircraft and spacecraft, it unfortunately presents a serious challenge for communication purposes.
This is because the introduction of the information signal itself introduces deviations from the equilibrium and the adaptation mechanism by its nature reacts against the information signal in an attempt to return to the equilibrium. This has the tendency of distorting the control signal at the receiver thereby resulting in a recovered signal that might be different to the original information. However, despite this challenge, we show that it is possible to recover the information signal at the receiver. In some cases, this is made possible by additional techniques such as filtering, exponential smoothing and design of special detection mechanism. In this report, cycle and bit accurate communication systems based on a new chaotic system are realized. The modulating signals are binary-valued bitstream and higher resolution signals of various sampling rates and frequencies. Applying a filtering or a specialized detection mechanism is dependent on the nature of the modulating signal. Furthermore, the designs are subjected to the real situation whereby channel noise is present. Despite transmission via an \textit{awgn} channel, most of the designs proved to be robust provided the \textit{SNR} is above a certain threshold or the noise power is below a certain threshold. 
\end{multicols}
\begin{multicols}{2}
\bibliographystyle{ieeetr}
\bibliography{securecomm}
\end{multicols}
\end{document}